\documentclass[a4paper,11pt]{iopart}
\usepackage{cite}
\usepackage{iopams}
\usepackage{tikz}
\usepackage{subfigure}
\usepackage{physics}

\begin{document}

\title{Current fluctuations in periodically driven systems}
\author{Andre C Barato$^1$ and Raphael Chetrite$^{2}$}
\address{
$^1$ Max Planck Institute for the Physics of Complex Systems,\\
N\"othnitzer Str. 38, 01187 Dresden, Germany\\
$^2$Laboratoire J A Dieudonn\'e,\\
UMR CNRS 7351, Universit\'e de Nice Sophia Antipolis, Nice 06108, France\\
}

\def\ex#1{\langle #1 \rangle}
\begin{abstract}
Small nonequelibrium systems driven by an external periodic protocol can be described
by Markov processes with time-periodic transition rates. In general, current fluctuations 
in such small systems are large and may play a crucial role.  We develop a 
theoretical formalism to evaluate the rate of such large deviations in periodically driven systems. We show
that the scaled cumulant generating function that  characterizes current fluctuations is given 
by a maximal Floquet exponent. Comparing deterministic protocols with stochastic protocols, 
we show that, with respect to large deviations, systems driven by a stochastic protocol with an infinitely large 
number of jumps are equivalent to systems driven by deterministic protocols. 
Our results are illustrated with three case studies: a two-state model for a heat engine,
a three-state model for a molecular pump, and a biased random walk with a time-periodic affinity.

\end{abstract}

%%%%%%%%%%%%%%%%%%%%%%%%%%%%%%%%%%%%%%%%%%%%%%%%%%%%%%%%%%%%%%%%%%%%%%%%%%%%%%%%%%%%%%%%%%%%%%%%%%%%%%%%%%%%%%%%%%%%%%%%%%%%%%%%%%%%%%%%%%%%%%%%%%%%%%%%%
\section{Introduction}
%%%%%%%%%%%%%%%%%%%%%%%%%%%%%%%%%%%%%%%%%%%%%%%%%%%%%%%%%%%%%%%%%%%%%%%%%%%%%%%%%%%%%%%%%%%%%%%%%%%%%%%%%%%%%%%%%%%%%%%%%%%%%%%%%%%%%%%%%%%%%%%%%%%%%%%%%

Periodic external control is used to operate a wide variety of thermodynamic machines that includes 
traditional idealized engines. Modern experimental examples of such machines are molecular pumps \cite{erba15} 
and micro-sized heat engines \cite{mart17}. For these small systems, thermodynamic currents, such as 
the work exerted on a molecular pump or the heat flow in a heat engine, display thermal fluctuations 
that can be relatively large. For instance, the prominent fluctuation theorem is a symmetry related 
to these fluctuations. 

Stochastic thermodynamics \cite{seif12} is an emerging field  that applies to small systems with 
large fluctuations. Within this theory, periodically driven systems are modeled as Markov processes with time-dependent
transition rates that are periodic. Such modeling  has been used in several works that include:
models for stochastic resonance \cite{gamm98}, linear response theory for periodically driven systems \cite{izum09,izum10,izum15,bran15,proe15,proe16,ray17}, 
theoretical studies for small heat engines far from equilibrium \cite{schm08,espo10,tu14,raz16a}, necessary conditions for 
the generation of a current in a molecular pump \cite{raha08,cher08,maes10,raha11,mand14}, a mapping relating periodically driven systems 
with systems driven by a fixed thermodynamic affinity \cite{raz16,rots17}, and 2.5 large deviations for Markov process with time-periodic generators \cite{sing10,bert18}.
However, a generic formalism to evaluate the rate of large deviations of single 
currents is still not available.

Periodically driven systems reach a limiting periodic state that can be contrasted with nonequilibrium 
stationary states, which are described by Markov  processes with constant transition rates. Physically, this second 
case corresponds to a system driven by a fixed thermodynamic affinity. For nonequilibrium 
stationary states, a generic formalism to quantify large deviations of currents is available: The so called scaled cumulant generating function 
(SCGF) is determined by the maximum eigenvalue of a tilted generator \cite{lebo99,koza99}. This formalism can be used to 
calculate the SCGF of systems subjected to a potential that is periodic in space \cite{spec12}, which is not the case for a 
time-periodic potential. 

In this paper, we develop a formalism to determine large deviations in periodically driven systems. 
We show that a fundamental (or monodromy) matrix from Floquet theory \cite{draz92,grim90,grif98,klau08},  which is related to a time-dependent 
tilted generator, quantifies current fluctuations.  Specifically, we show that this fundamental matrix 
is a Perron-Frobenius matrix and that its maximal eigenvalue gives the SCGF.

Deterministic protocols are commonly used in the study of periodically driven systems. Nevertheless, such systems can also be driven by a cyclic stochastic 
protocol that mimics the periodicity of a deterministic protocol \cite{ray17,verl14,bara16,bara17b}. In this case, the system and protocol together form a bipartite Markov process with time-independent 
transition rates. For stochastic protocols, current fluctuations can then be analyzed within the stationary state of this 
bipartite Markov process. We prove the equivalence of current fluctuations between systems driven by a 
deterministic protocol and systems driven by a stochastic protocol with an infinitely large number of jumps. Therefore, we show that 
a periodic protocol can be seen as a particular limit of a stochastic protocol.  

Illustrations of our results are performed with three models. An exactly solvable model 
for a heat engine, a model for molecular pump that we use to compare theory with numerical simulations, and 
one model for a biased random walk with a time-periodic affinity that has a particularly simple 
expression for the SCGF.

The paper is organized in the following way. In Section \ref{sec2}, we define fluctuating currents in Markov processes with time-periodic transition rates. 
The formalism for the calculation of the SCGF is developed in Section \ref{sec3}. In Section \ref{sec4}, we analyze
three case studies. Section \ref{sec5} contains the results for a stochastic protocol. We conclude in Section \ref{sec6}. 
In \ref{appnew}, we discuss fluctuating currents such as work, which can be written in a form that is apparently different 
from the generic currents we consider in the main text. We extend our results to diffusion processes in \ref{appa}. 
An exact calculation of the SCGF for systems with two states and a piecewise constant protocol is presented in \ref{appb}.    

%%%%%%%%%%%%%%%%%%%%%%%%%%%%%%%%%%%%%%%%%%%%%%%%%%%%%%%%%%%%%%%%%%%%%%%%%%%%%%%%%%%%%%%%%%%%%%%%%%%%%%%%%%%%%%%%%%%%%%%%%%%%%%%%%%%%%%%%%%%%%%%%%%%%%%%%%
\section{General setup and mathematical definitions}
%%%%%%%%%%%%%%%%%%%%%%%%%%%%%%%%%%%%%%%%%%%%%%%%%%%%%%%%%%%%%%%%%%%%%%%%%%%%%%%%%%%%%%%%%%%%%%%%%%%%%%%%%%%%%%%%%%%%%%%%%%%%%%%%%%%%%%%%%%%%%%%%%%%%%%%%%
\label{sec2}

We consider a Markov process with a finite number of states 
$\Omega$. The time-dependent transition rate from state $i$
to state $j$ at time $t$ is denoted by $w_{ij}(t)$. These transition rates are 
time-periodic with a period $\tau$, i.e., 
\begin{equation}
w_{ij}(t+\tau)= w_{ij}(t).
\end{equation}
In the theoretical framework of stochastic thermodynamics, which we use in our illustrative examples, the transition rates 
fulfill the restriction that if $w_{ij}(t)\neq 0$ then $w_{ji}(t)\neq 0$. However, our mathematical results (in Section \ref{sec3} and Section \ref{sec5})
do not rely on this assumption.  A physical interpretation for these transition rates based on the generalized detailed balance relation from stochastic thermodynamics 
can be found in \cite{bran15,proe16,ray17}.

The associated master equation reads 
\begin{equation}
\frac{d}{dt} P(i,t)= \sum_j\left[P(j,t)w_{ji}(t)-P(i,t)w_{ij}(t)\right],
\label{meq}
\end{equation}
where $P(i,t)$ is the probability to be in state $i$ at time $t$. 
This equation can be written in the vectorial form 
\begin{equation}
\frac{d}{dt} \ket{P_t}= \mathcal{L}_t\ket{P_t},
\end{equation}
where $\ket{P_t}$ is a vector with components $P(i,t)$  and $\mathcal{L}_t$ is the stochastic matrix defined as 
\begin{equation}
[\mathcal{L}_t]_{ji}\equiv (1-\delta_{ij})w_{ij}(t) -\delta_{ij}\sum_kw_{ik}(t). 
\end{equation}

A stochastic trajectory from time $0$ to time $T=n\tau$ is a sequence of jumps and waiting times, which is denoted by $A_0^T$. If a jump takes place at time $t$, the state before the jump 
is denoted $a_t^{-}$ and the state after the jump is denoted $a_t^{+}$. A fluctuating current is a functional of the stochastic trajectory defined as 
\begin{equation}
X[A_0^T]\equiv\sum_{0\le t\le T} \theta_{a_t^{-},a_t^{+}}(t).
\label{currtraj}
\end{equation}
For a current, the increments $\theta_{i,j}(t)$ are anti-symmetric, i.e., 
\begin{equation}
\theta_{i,j}(t)=-\theta_{j,i}(t). 
\label{currincre}
\end{equation}
Fluctuations of this current in the long time limit are characterized by  the SCGF
\begin{equation}
\lambda(z)\equiv\lim_{T\to\infty}\frac{1}{T}\ln\langle\textrm{e}^{zX}\rangle=\frac{1}{\tau}\lim_{n\to\infty}\frac{1}{n}\ln\langle\textrm{e}^{zX}\rangle,
\label{scaleddef}
\end{equation}
where the brackets mean an average over stochastic trajectories. 
The average current $J$ and diffusion coefficient $D$ are given by 
\begin{equation}
J\equiv \lim_{T\to\infty}\frac{1}{T}\langle X\rangle= \lambda'(0)
\label{eqJ}
\end{equation}
and
\begin{equation}
D \equiv \lim_{T\to\infty}\frac{1}{T}\left(\langle X^2\rangle-\langle X\rangle^2\right)= \lambda''(0),
\label{eqdif}
\end{equation}
respectively. Similarly, higher order moments associated with $X$ can be obtained by taking higher order derivatives
of $\lambda(z)$ at $z=0$.  

In the long time limit, the system reaches an invariant limiting periodic distribution $P^{\textrm{inv}}_i(t)=P^{\textrm{inv}}_i(t+\tau)$.
The average current can be written in terms of this distribution as
\begin{equation}
J=  \frac{1}{\tau}\int_0^\tau\sum_{i<j}\theta_{ij}(t)\left[P^{\textrm{inv}}_i(t)w_{ij}(t)-P^{\textrm{inv}}_j(t)w_{ji}(t)\right]dt,
\end{equation}
where $\sum_{i<j}$ means a sum over all links with non-zero transition rates in the network of states. 

An important observable in stochastic thermodynamics is the work done on the system due to the periodic variation of 
energy of the system. This fluctuating current is typically written as an integral of a function over the time interval $T$. However, 
as we show in appendix \ref{appnew}, observables such as work can also be written in the form given by Eq. \eqref{currtraj}.

%%%%%%%%%%%%%%%%%%%%%%%%%%%%%%%%%%%%%%%%%%%%%%%%%%%%%%%%%%%%%%%%%%%%%%%%%%%%%%%%%%%%%%%%%%%%%%%%%%%%%%%%%%%%%%%%%%%%%%%%%%%%%%%%%%%%%%%%%%%%%%%%%%%%%%%%%
\section{Floquet theory for the SCGF}
%%%%%%%%%%%%%%%%%%%%%%%%%%%%%%%%%%%%%%%%%%%%%%%%%%%%%%%%%%%%%%%%%%%%%%%%%%%%%%%%%%%%%%%%%%%%%%%%%%%%%%%%%%%%%%%%%%%%%%%%%%%%%%%%%%%%%%%%%%%%%%%%%%%%%%%%%
\label{sec3}

\subsection{General theory}

The joint probability that the current is $X$ and the system is in state $i$ at time $t$ is written as $P(i,X,t)$, whereas the vector 
$\ket{P_t(X)}$ has components $P(i,X,t)$. The Laplace transform of $\ket{P_t(X)}$ is given by 
\begin{equation}
\ket{\tilde{P}_t(z)}\equiv\sum_X \textrm{e}^{Xz} \ket{P_t(X)}.
\label{eqlaplace}
\end{equation}
The average $\langle\textrm{e}^{zX}\rangle$ in Eq. \eqref{scaleddef} is related to this Laplace transform in the following way,  
\begin{equation}
\langle\textrm{e}^{zX}\rangle= \sum_{i=1}^{\Omega} \tilde{P}(i,z,T).
\label{scaledp}
\end{equation}
From the master equation \eqref{meq} and the tilted generator 
\begin{equation}
[\mathcal{L}_t(z)]_{ji}\equiv (1-\delta_{ij})w_{ij}(t)\textrm{e}^{z\theta_{ij}t} -\delta_{ij}\sum_kw_{ik}(t),
\label{defL}
\end{equation}
we obtain
\begin{equation}
\frac{d}{dt}\ket{\tilde{P}_t(z)}=\mathcal{L}_t(z)\ket{\tilde{P}_t(z)}.
\label{laplaceevo}
\end{equation}
Using the periodicity of $\mathcal{L}_t(z)$, the formal solution of this equation at time $T=n\tau$ is
\begin{equation}
\ket{\tilde{P}_T(z)}=\overleftarrow{\exp}\left(\int_0^T\mathcal{L}_t(z)dt\right) \ket{\tilde{P}_0(z)}= \left[\overleftarrow{\exp}\left(\int_0^\tau\mathcal{L}_t(z)dt\right)\right]^{n} \ket{\tilde{P}_0(z)}\equiv \mathcal{M}(z)^n\ket{\tilde{P}_0(z)},
\label{Mdef}
\end{equation}
where $\ket{\tilde{P}_0(z)}$ is the initial condition and $\overleftarrow{\exp}$ represents a time-reversed ordered exponential. 
This ordered exponential can be defined as the solution of the differential equation  
\begin{equation}
\frac{d}{dt} \overleftarrow{\exp}\left(\int_0^t\mathcal{L}_{t'}(z)dt'\right)= \mathcal{L}_t(z)\overleftarrow{\exp}\left(\int_0^t\mathcal{L}_{t'}(z)dt'\right),
\label{numeq}
\end{equation}
where the initial condition is the identity matrix. 

The matrix 
\begin{equation}
\mathcal{M}(z)\equiv \overleftarrow{\exp}\left(\int_0^\tau\mathcal{L}_{t'}(z)dt'\right)
\label{Mdefdef}
\end{equation}
is a central object that is known as fundamental matrix in Floquet theory \cite{klau08}. 
The eigenvalues of this matrix are denoted by $\rho_k(z)$, the right  eigenvectors by $\ket{r_k(z)}$, and 
the left eigenvectors by $\bra{l_k(z)}$. The fundamental matrix can then be written as 
\begin{equation}
\mathcal{M}(z)=   \sum_{k=1}^{\Omega}\rho_k(z)\ket{r_k(z)}\bra{l_k(z)}.
\label{mexp}
\end{equation}
From Eq. \eqref{Mdef}, by Setting $T=\tau$, imposing the initial condition $a_0=i$, and 
restricting to trajectories that finish at state $a_T=j$, we obtain  
\begin{equation}
\langle\textrm{e}^{zX}\delta_{a_T,j}|a_0=i\rangle= [\mathcal{M}(z)]_{ji}.
\label{Mzpositive}
\end{equation}
This equation shows that all elements of the fundamental matrix $\mathcal{M}(z)$ are positive. Hence, from the Perron-Frobenius theorem, the matrix $\mathcal{M}(z)$ has 
a maximal real eigenvalue defined as  $\rho_1(z)$. This fact together with the definition of the SCGF
in Eq. \eqref{scaleddef}, Eq. \eqref{scaledp}, Eq.  \eqref{Mdef},  and Eq.  \eqref{mexp} 
lead to the main result 
\begin{equation}
\lambda(z)\simeq \frac{1}{\tau}\ln \rho_1(z),  
\label{eqmainresult}  
\end{equation}
where the symbol $\simeq$ means asymptotic equality in the limit $n\to \infty$ and $\tau^{-1}\ln \rho_1(z)$ is the maximal Floquet exponent. 
The SCGF $\lambda(z)$ can be evaluated by first 
solving Eq. \eqref{numeq} and then calculating the maximal eigenvalue of $\mathcal{M}(z)$. 
For $z=0$ the matrix elements in Eq. \eqref{Mzpositive} are transition probabilities, therefore, 
the maximum eigenvalue of the matrix is  $\rho_1(0)=1$, which implies $\lambda(0)=0$. It is worth noting that 
Eq. \eqref{eqmainresult} is not restricted to currents. This result is also valid 
for any functional of the stochastic trajectory with the form given in Eq. \eqref{currtraj} that has increments that do not
fulfill the anti-symmetry in Eq. \eqref{currincre}. We point out that 
the SCGF has been obtained as a maximum Floquet exponent for specific two-state models in \cite{ray17,verl13}.

This relation between SCGF and maximal Floquet exponent is also valid for diffusion processes, as shown in \ref{appa}. 
For piecewise protocols, transition rates $w_{ij}(t)$ are piecewise. 
If the external protocol is piecewise constant, with the period 
divided into $L$ pieces and $\tau=\tau_0+\tau_1+\ldots+\tau_{L-1}$, the matrix  $\mathcal{M}(z)$ 
defined in Eq. \eqref{numeq} takes the form 
\begin{equation}
\mathcal{M}(z)= \exp(\tilde{\mathcal{L}}_{L-1}(z)\tau_{L-1})\ldots\exp(\tilde{\mathcal{L}}_1(z)\tau_{1})\exp(\tilde{\mathcal{L}}_{0}(z)\tau_{0}),  
\label{Mpiece}    
\end{equation}
where $\tilde{\mathcal{L}}_{k}(z)$ is the constant modified generator during the interval $\tau_k$. In this last equation, it is assumed that 
the increments of the current are also piecewise constant. The SCGF
can then be obtained from the maximal eigenvalue of this matrix. For a piecewise protocol that is 
not constant, the expression of $\mathcal{M}(z)$ becomes a product of ordered exponentials.

\subsection{Expressions for average current and diffusion coefficient}  

The average current $J$, diffusion coefficient $D$, and higher order cumulants can be obtained without explicit 
evaluation of the maximum eigenvalue associated with $\mathcal{M}(z)$ in the following way. A similar method for non-equilibrium stationary states 
has been introduced by Koza \cite{koza99} (see also \cite{bara15,wach15}). The characteristic polynomial associated with $\mathcal{M}(z)$ is written as
\begin{equation}
\det\left(x\mathbf{I}-\mathcal{M}(z)\right)=\sum_{m=0}^{\Omega}c_{m}(z)x^{m},
\end{equation}
where $\mathbf{I}$ is the identity matrix.
The maximum eigenvalue $\rho_1(z)$ is a root of this polynomial, which leads to the equation
\begin{equation}
\sum_{m=0}^{\Omega}c_{m}(z)\left[\rho_{1}(z)\right]^{m}=0.
\label{eqchara}
\end{equation}
Taking a derivative with respect to $z$ and setting $z=0$ we obtain 
\begin{equation}
J= \frac{1}{\tau}\frac{\rho'_{1}(0)}{\rho_{1}(0)}=-\frac{1}{\tau}\frac{\sum_{m=0}^{\Omega}c'_{m}(0)}{\sum_{m=0}^{\Omega}mc_{m}(0)},
\label{eqkozaj}
\end{equation}
where we used Eq. \eqref{eqJ} and $\rho_{1}(0)=1$. Taking a 
second derivative with respect to $z$ of Eq. \eqref{eqchara} and setting $z=0$ we obtain  
\begin{align}
D & = \frac{1}{\tau}\left\{\rho''_{1}(0)-[\rho'_{1}(0)]^2\right\}\nonumber\\
 & =-\frac{\sum_{m=0}^{\Omega}c''_{m}(0)+2\rho'_{1}(0)\sum_{m=0}^{\Omega}mc'_{m}(0)+\left[\rho'_{1}(0)\right]^{2}\sum_{m=0}^{\Omega}m^2c_{m}(0)}{\tau\sum_{m=0}^{\Omega}mc_{m}(0)},
\label{eqkozaD}
\end{align}
where we used Eq. \eqref{eqdif} and $\rho_{1}(0)=1$. Using these expressions,  $J$ and $D$ can be evaluated directly from the coefficients of the characteristic polynomial associated with 
$\mathcal{M}(z)$. Taking higher order derivatives lead to similar expressions for higher order cumulants.

%\begin{figure}
%\centering
%\input{fig3a}
%\input{fig3b}
%\includegraphics[width=60mm]{./0k7tau.eps}
%\includegraphics[width=62mm]{./0tauk2.eps}
%\caption{Lower and upper bounds obtained from (\ref{bounds}) for the discrete time version of the model of Fig. \ref{fig1}. The parameters are $\gamma=1$, $B_1=0$, $B_2=\ln(10)$,
%$k=7$ (left panel), and $k=2$ (right panel). In the limit $\tau\to 0$, the upper bounds go to the value given by (\ref{ubi}) and the lower bounds go to zero.} 
%\label{fig4}
%\end{figure}

%%%%%%%%%%%%%%%%%%%%%%%%%%%%%%%%%%%%%%%%%%%%%%%%%%%%%%%%%%%%%%%%%%%%%%%%%%%%%%%%%%%%%%%%%%%%%%%%%%%%%%%%%%%%%%%%%%%%%%%%%%%%%%%%%%%%%%%%%%%%%%%%%%%%%%%%%
\section{Case studies}
%%%%%%%%%%%%%%%%%%%%%%%%%%%%%%%%%%%%%%%%%%%%%%%%%%%%%%%%%%%%%%%%%%%%%%%%%%%%%%%%%%%%%%%%%%%%%%%%%%%%%%%%%%%%%%%%%%%%%%%%%%%%%%%%%%%%%%%%%%%%%%%%%%%%%%%%%
\label{sec4}

\subsection{Heat engine}
\label{sec41}

\begin{figure}
\subfigure[]{\includegraphics[width=60mm]{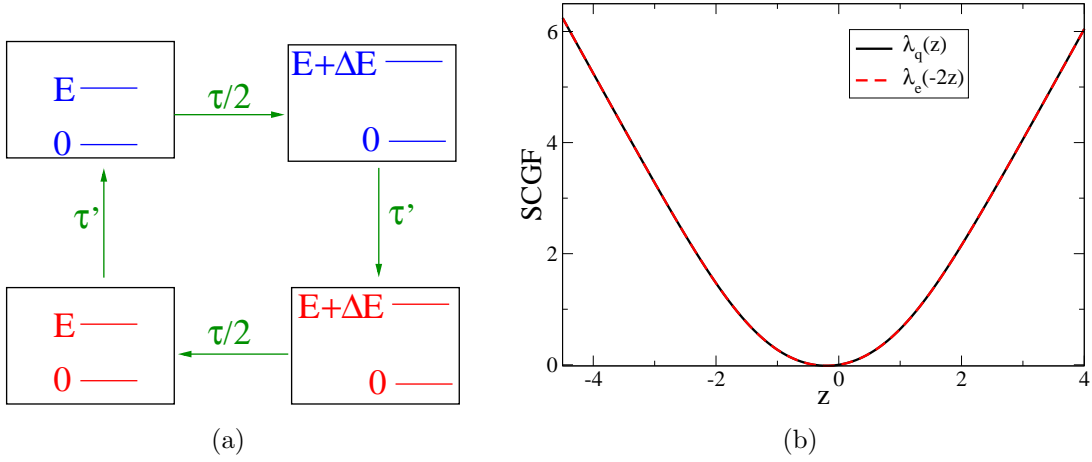}\label{fig1a}}
\hspace{5mm}
\subfigure[]{\includegraphics[width=75mm]{fig1b.eps}\label{fig1b}}
\vspace{-3mm}
\caption{Model and results for a heat engine. (a) Illustration of the model. The energy levels in blue (red) are associated with the cold (hot) temperature 
$\beta_c^{-1}$ ($\beta_h^{-1}$). (b) SCGF associated with the heat current 
$\lambda_q(z)$ and the work current $\lambda_e(z)$. Parameters are set to $\Delta E=E=\beta_c=\tau=k=1$ and $\beta_h=1/10$, which 
gives $\beta_c(E+\Delta E)=2$. The heat and work currents are not independent in this model with their SCGF 
following the relation $\lambda_q(z)=\lambda_e(-\beta_c(E+\Delta E)z)$.
}
\label{fig1}
\end{figure}

We introduce an exactly solvable two-state model for a heat engine with a piecewise protocol, which is illustrated in Fig. \ref{fig1a}. This model is similar to a model for a heat 
engine with a stochastic protocol analyzed in \cite{ray17}. One of the states has energy 0 and the other state 
has a higher energy that depends on time. The protocol is divided in four steps. First, the energy changes from $E$ to $E+\Delta E$ at a cold temperature $\beta_c^{-1}$.
Second, the temperature changes from $\beta_c^{-1}$ to the hot temperature $\beta_h^{-1}$. Third, the energy changes back from $E+\Delta E$ to $E$. Fourth, the temperature 
changes back from $\beta_h^{-1}$ to $\beta_c^{-1}$. The inverse temperature takes the form 
\begin{equation}
\beta^k= \beta_c[1-F_qh^k],
\end{equation}
where $F_q\equiv(\beta_c-\beta_h)/\beta_c\ge 0$, $h^0=h^1=0$ and $h^2=h^3=1$. Physically, $F_q$ is the thermodynamic affinity associated with the heat current \cite{ray17}
and the temperature depends on time through the parameter $h^k$. The energy of the state with higher energy is given by   
\begin{equation}
E^k= E+\Delta E f^k,
\end{equation}
where $f^0=f^3=0$, $f^1=f^2=1$. The parameter $f^k$ gives the time-dependence of the energy and $\Delta E$ is the amplitude of the time-dependent part of the energy.

The time-intervals of the period are set to $\tau_0=\tau_2=\tau/2$ and $\tau_1=\tau_3=\tau'\to 0$, i.e.,
the energy changes happens after a time interval $\tau/2$ and the temperature changes are instantaneous.
Hence, the number of pieces of the protocol is reduced from four to two. The transition rates for this model, which fulfill 
the generalized detailed balance relation \cite{seif12}, are set to 
\begin{equation}
w_+^0=w\textrm{e}^{-\beta_c E/2}\qquad\textrm{and}\qquad  w_-^0=w\textrm{e}^{\beta_c E/2}, 
\end{equation}
for the first part of the period; 
\begin{equation}
w_+^1=w\textrm{e}^{-\beta_h(E+\Delta E)/2}\qquad\textrm{and}\qquad  w_-^1=w\textrm{e}^{\beta_h (E+\Delta E)/2},
\end{equation}
for the second part of the period.
The subscript $+$ indicates a transition rate from the state with energy zero to the state with energy $E^k$, whereas the subscript $-$ indicates the reversed  transition rate.
The superscript $0$ indicates the first half of the the period and the superscript $1$ indicates the second half of the period. 
  
The basic physics of the model is that part of the heat taken from the hot reservoir is transformed into extracted work, as explained in \cite{ray17}.
Two currents of interest are the heat current $X_q$ and the work current $X_e$. 
The piecewise version of the modified generator for a generic current reads  
\begin{equation}
\tilde{\mathcal{L}}_{k}(z)=\left(\begin{array}{cc}
-w_+^{k} & w_-^{k}\textrm{e}^{-z\theta^k}\\
w_+^{k}\textrm{e}^{z\theta^k} & -w_-^{k}
\end{array}\right),
\end{equation}
where $k=0$ for the first half of the period and $k=1$ for the second half of the period.
The increments for the heat current $X_q$ are defined as $\theta_q^0\equiv 0$ and $\theta_q^1\equiv\beta_c(E+\Delta E)$. The increments for the work current $X_e$ 
are defined as $\theta_e^0\equiv 0$ and $\theta_e^1\equiv-1$. 
Hence, we obtain the relation $X_q=-\beta_c(E+\Delta E)X_e$, which with Eq. \eqref{scaleddef} leads to 
$\lambda_q\left(z\right)=\lambda_e\left(-\beta_c(E+\Delta E)z\right)$ for the SCGF. In other words, for this simple model there is tight coupling
between the work and heat currents, as shown in Fig. \ref{fig1b}. 

The exact calculation of the SCGF for a generic piecewise two-state model, is presented in  \ref{appb}. 
For the present model, the SCGF associated with $X_q$ is
\begin{equation}
\lambda_q(z)=\frac{1}{\tau}\ln\left[\frac{f(z)+\sqrt{\left[f(z)\right]^{2}-4\textrm{e}^{-(w^0+w^1)/2}}}{2}\right]
\end{equation}
where
\begin{align}
f(z)\equiv & 4\textrm{e}^{-\tau(w^0+w^1)/4}\sinh(w^0\tau/4)\sinh(w^1\tau/4)[q_+^1q_-^0\textrm{e}^{z\beta_c(E+\Delta E)}+q_-^1q_+^0\textrm{e}^{-z\beta_c(E+\Delta E)}]\nonumber\\
&+(\textrm{e}^{-\tau w^0/2}q_+^0+q_-^0)(\textrm{e}^{-\tau w^1/2}q_+^1+q_-^1)+(q_+^0+\textrm{e}^{-\tau w^0/2}q_-^0)(q_+^1+q_-^1\textrm{e}^{-\tau w^1/2}),
\end{align}
$w^0\equiv 2w \cosh(\beta_c E/2)$, $w^1\equiv 2w \cosh(\beta_h(E+\Delta E)/2)$, $q_\pm^0\equiv w_\pm^0/w^0$, and $q_\pm^1\equiv w_\pm^1/w^1$.
The SCGF plotted  in Fig. \ref{fig1b}  is a concave function of $z$, which is a generic property of a SCGF, and it becomes linear 
in $z$ for large $z$, which is a peculiarity of two-state models. 

\subsection{Molecular pump}
\label{sec42}

We now analyze a three-state model for a molecular pump with states $i=1,2,3$, which is similar to a model analyzed in \cite{raha08}.
The key phenomena that happens in such pumps, is that even though there are no fixed thermodynamic affinities, a suitable time-periodic 
variation of energies $E_i(t)$ and energy barriers $B_{i}(t)$ can lead to net rotation in the three state system. Periodically driven 
molecular pumps can be realized experimentally with interlocked molecular rings \cite{erba15}. The transition rates are set to   
\begin{equation}
w_{ii+1}=w\textrm{e}^{E_i(t)-B_{i+1}(t)}\qquad\textrm{and}\qquad w_{i+1i}=w\textrm{e}^{E_{i+1}(t)-B_{i+1}(t)},
\end{equation}
where for $i=3$ we have $i+1=1$ and the inverse temperature is $\beta=1$. The energies are given by 
\begin{equation}
E_i(t)= -1+\cos[2\pi (t+(i-1)/3)/\tau],
\end{equation}
and the energy barriers are given by 
\begin{equation}
B_i(t)= \cos[2\pi (t+(i-1)/3)/\tau].
\end{equation}

\begin{figure}
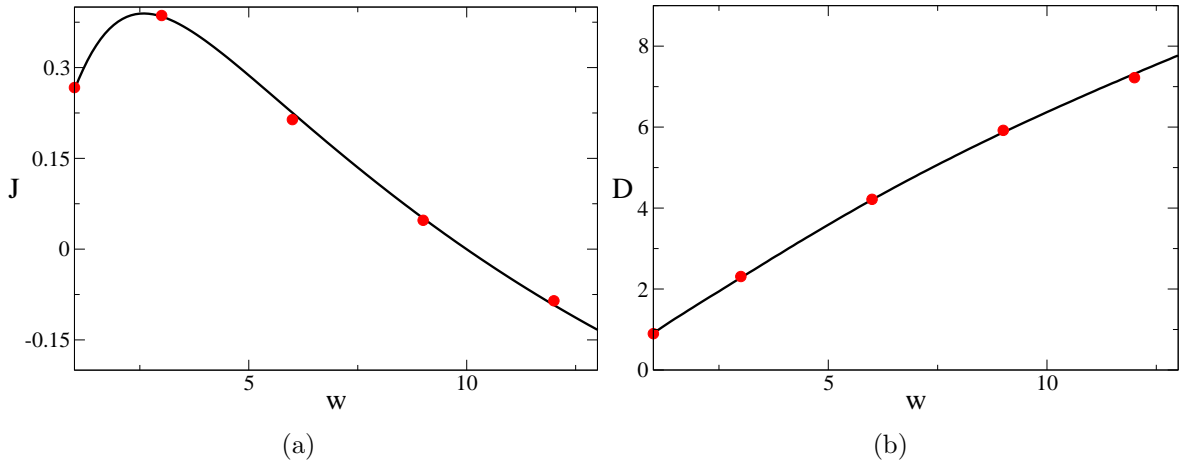

\subfigure[]{\includegraphics[width=78mm]{fig2a.eps}\label{fig2a}}
\subfigure[]{\includegraphics[width=75mm]{fig2b.eps}\label{fig2b}}
\vspace{-3mm}
\caption{Average current $J$ and diffusion coefficient $D$ for the molecular pump. The black line 
was obtained from the numerical evaluation of $\mathcal{M}(z)$ and the red dots from Monte-Carlo simulations. The period is set to $\tau=1$.   
}
\label{fig2}
\end{figure}

The increment of the fluctuating current of interest is such that it increases by one if there is a transition in the clockwise direction
($1\to 2$, $2\to 3$, and $3\to 1$) and it decreases by one if there is a transition in the counter clockwise direction.
The time-dependent modified generator for this current reads  
\begin{equation}
\mathcal{L}_{t}(z)=\left(\begin{array}{ccc}
 -w_{12}(t) - w_{13}(t) & w_{21}(t)\textrm{e}^{-z} & w_{31}(t)\textrm{e}^{z}\\
 w_{12}(t)\textrm{e}^{z} &  -w_{21}(t) - w_{23}(t) & w_{32}(t)\textrm{e}^{-z}\\
 w_{13}(t)\textrm{e}^{-z} & w_{23}(t)\textrm{e}^{z} & -w_{31}(t) - w_{32}(t)
\end{array}\right).
\label{modifiedpump}
\end{equation}
We calculated the average current $J$ and diffusion coefficient $D$ using Eq. \eqref{eqkozaj} and Eq. \eqref{eqkozaD}, respectively.  
The fundamental matrix $\mathcal{M}(z)$ was evaluated with numerical solution of Eq. \eqref{numeq}. In Fig. \ref{fig2} we show that both quantities  show perfect 
agreement with results obtained from Monte Carlo simulations. Note that the sign of the current depends on the rate constant $w$ that determines 
the speed of the transitions in relation to the period $\tau$: for large enough $w$ the current becomes negative. We have used a 
discrete-time algorithm for our simulations with a sufficiently small time-step,
a continuous-time algorithm for time-dependent transition rates can be found in \cite{prad97}.

\subsection{Biased random walk with time-periodic affinity}
\label{sec43}

In this Section, we consider a biased random walk on a ring with $\Omega$ states  driven by a time-periodic affinity 
\begin{equation}
F(t)\equiv F \cos(2\pi t/\tau).
\end{equation}
The transition rate for a jump in the clockwise direction is given by 
\begin{equation}
w_+(t)= w\exp[\psi F(t)/\Omega],
\end{equation}
whereas the transition rate for 
a jump in the anti-clockwise direction is 
\begin{equation}
w_-(t)= w\exp[(\psi -1)F(t)/\Omega],
\end{equation}
where $0\le \psi\le 1$. The parameter $\psi$ determines how $F(t)$  
influences forward and backward rates. We consider the current in the ring, which increases by one if a jump in the clockwise 
direction takes place and decreases by one if a jump in the anti-clockwise direction takes place. For example, for $\Omega=3$ the modified 
generator   $\mathcal{L}_{t}(z)$ takes the form in Eq. \eqref{modifiedpump} with $w_{12}(t)=w_{23}(t)=w_{31}(t)= w_+(t)$ and 
$w_{21}(t)=w_{32}(t)=w_{13}(t)= w_-(t)$.

This model has the peculiar property that the uniform vector $\bra{1}$ is a left eigenvector of $\mathcal{L}_{t}(z)$, for all $t$ and $z$, with eigenvalue 
\begin{equation}
\lambda_t(z)= -w_-(t)-w_+(t)+w_-(t)\textrm{e}^{-z}+w_+(t)\textrm{e}^{z}.
\label{lambdaRW}
\end{equation}
Hence, from the Dyson series of the ordered exponential and  Eq. \eqref{Mdefdef} for $\mathcal{M}(z)$, we obtain
\begin{equation}
\bra{1}\mathcal{M}(z)= \bra{1}\exp\left(\int_0^\tau\lambda_t(z)dt\right).
\label{Mdef2}
\end{equation}
Since $\bra{1}$ is a positive vector, from the Perron-Forbenius  theorem, it must be the eigenvector associated with the maximum eigenvalue of $\mathcal{M}(z)$, 
which leads to $\rho_1(z)=\exp\left(\int_0^\tau\lambda_t(z)dt\right)$. Using Eq. \eqref{eqmainresult}, we then obtain  
\begin{equation}
\lambda(z)= \tau^{-1}\int_0^\tau\lambda_t(z)dt.
\label{Mdef3}
\end{equation}
Explicit evaluation of the above integral leads to 
\begin{equation}
\lambda(z)= w(1-\textrm{e}^{-z}) [\textrm{e}^{z}I_0(\psi F_0/\Omega)-I_0( F_0(-1+\psi)/\Omega)],
\label{Mdef4}
\end{equation}
where $I_0(x)$ is a modified Bessel function of the first kind. Interestingly, the average current in this model is given by 
\begin{equation}
J=\lambda'(0)= w[I_0(\psi F/\Omega)-I_0( F(-1+\psi)/\Omega)].
\label{Mdef5}
\end{equation}
Even though the thermodynamic affinity integrated over a period is zero, the average current can be non-zero: $J$ is 
positive for $\psi>1/2$ and negative for $\psi<1/2$.

\begin{figure}
\centering\includegraphics[width=78mm]{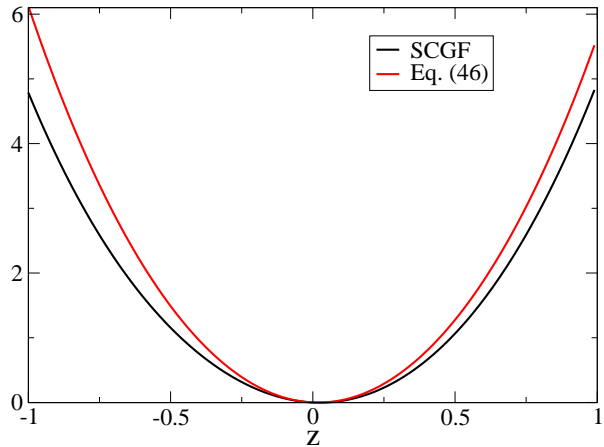}
\vspace{-3mm}
\caption{Comparison between SCGF obtained with the formalism from Section \ref{sec3} and 
the right hand side of Eq. \eqref{peculiarcond2}. These results were obtained  
for the model from Section \ref{sec42} with $\tau= 1$ and $w=5$.
}
\label{fig3}
\end{figure}

The expression in Eq. \eqref{Mdef3} for $\lambda(z)$ in terms of an integral of the maximum eigenvalue associated 
with $\mathcal{L}_{t}(z)$ is valid not only for the above model but for any current related to a modified generator that 
fulfils the property  
\begin{equation}
\bra{1}\mathcal{L}_{t}(z)= \bra{1}\lambda_t(z),
\label{peculiarcond}
\end{equation}
for all $z$ and $t$. Furthermore, from Eq. \eqref{Mdef3}, if the property \eqref{peculiarcond} is satisfied, the SCGF can be written as 
\begin{equation} 
\lambda(z)= \tau^{-1}\sum_{ij}\int_0^\tau P^{\textrm{inv}}_i(t)[\mathcal{L}_{t}(z)]_{ji}dt,
\label{peculiarcond2}
\end{equation}
where this equation is valid not only for $P^{\textrm{inv}}_i(t)$ but for an arbitrary probability distribution.
Expression \eqref{peculiarcond2} for the evaluation of the SCGF has been proposed in \cite{rots17} as a general expression for the SCGF. 
While it is correct for this peculiar case, in general, Eq. \eqref{peculiarcond2} does not provide the correct SCGF. In Fig. \ref{fig3}, we show that 
the right hand side of Eq. \eqref{peculiarcond2} is different from the SCGF for the model for a molecular pump from 
Section \ref{sec42}.

%%%%%%%%%%%%%%%%%%%%%%%%%%%%%%%%%%%%%%%%%%%%%%%%%%%%%%%%%%%%%%%%%%%%%%%%%%%%%%%%%%%%%%%%%%%%%%%%%%%%%%%%%%%%%%%%%%%%%%%%%%%%%%%%%%%%%%%%%%%%%%%%%%%%%%%%%
\section{Stochastic Protocol}
%%%%%%%%%%%%%%%%%%%%%%%%%%%%%%%%%%%%%%%%%%%%%%%%%%%%%%%%%%%%%%%%%%%%%%%%%%%%%%%%%%%%%%%%%%%%%%%%%%%%%%%%%%%%%%%%%%%%%%%%%%%%%%%%%%%%%%%%%%%%%%%%%%%%%%%%%
\label{sec5}

Hitherto we have restricted to the case of a deterministic protocol. In this section, we consider a system driven by a stochastic
protocol, which is cyclic and, therefore, mimics periodicity. We show that a stochastic protocol with a infinitely large number of jumps is 
equivalent to a deterministic protocol with respect to the large deviations of fluctuating currents.

For a stochastic protocol, the mathematical model is a bipartite Markov process with time-independent transition rates \cite{bara16,ray17}.
The bipartite Markov process has $\Omega\times N$ states, where $N$ is the number of jumps of the external protocol. A state of the bipartite 
Markov process $(i,n)$ is determined by the variable $i$ that identifies the state of the system and the variable 
$n=0,1,\ldots,N-1$ that identifies the state of the external protocol. The transition rates 
for a change in the state of the system are defined as 
\begin{equation}
w_{ij}^n\equiv w_{ij}(t=n\tau/N),
\end{equation}
where $w_{ij}(t)$ is the transition rate of a corresponding Markov process with time-dependent transition rates that describes a 
deterministic protocol.  

Since the external protocol is stochastic, there is a transition rate associated with changes of the state of the protocol from 
$n$ to $n+1$. The reversed transition rate is zero. 
From the periodicity of the protocol, for $n=N-1$ the protocol transitions back to $n=0$. The transition rate for a change in the protocol 
is set to $N/\tau$, hence, the average time for the protocol to complete a cycle is $\tau$. 

The stationary distribution of state $(i,n)$ is denoted $P_i^n$, where the dependence on the total number of
jumps $N$ is not shown for a compact notation. The conditional probability of state $i$ given that the protocol 
is in state $n$ is $P(i|n)=P_i^n/P^n$, where $P^n\equiv\sum_iP_i^n=1/N$ is the stationary probability that the protocol is 
at state $n$. As shown in \cite{bara16}, in the limit of $N\to\infty$, the stationary distribution of the bipartite Markov process is equivalent 
to the invariant periodic distribution of the corresponding Markov process with time-periodic transition rates, i.e., 
$P(i|n)\to P^{\textrm{inv}}_i(t)$, where $\tau n/N\to t$. 

A stochastic trajectory  of the bipartite process, from time $0$ to time $T'$, is denoted by $(A,\Xi)_0^{T'}$, where $A$ represents the state of the system  
and $\Xi$ represents the state of the protocol. A generic current, analogous to the current in Eq. \eqref{currtraj} for a deterministic 
protocol, is defined as
\begin{equation}
X_N[(A,\Xi)_0^{T'}]\equiv\sum_{0\le t'\le T'} \theta_{a_t'^{-},a_t'^{+}}^{\xi_t'},
\end{equation}
where $\theta_{ij}^n\equiv \theta_{ij}(t=n\tau/N)$. It can be shown that the SCGF associated with this current 
can be obtained from the tilted generator  $\mathcal{L}(z)$ \cite{bara16}, which is a matrix with dimension $N\times\Omega$ given by
 \begin{equation}
\mathcal{L}(z)=\left(\begin{array}{cccccc}
\mathcal{L}_{0}(z)-\mathbf{I}N/\tau & 0 & 0 & \ldots &  \mathbf{I}N/\tau\\
\mathbf{I}N/\tau & \mathcal{L}_{\frac{\tau}{N}}(z)-\mathbf{I}N/\tau &  \ldots  & 0 & 0\\
0 & \mathbf{I}N/\tau &   \mathcal{L}_{\frac{2\tau}{N}}(z)-\mathbf{I}N/\tau &\ldots    & 0\\
\vdots &  \vdots & \vdots & \vdots &  \vdots \\
0 & 0 & 0 & \ldots  &   \mathcal{L}_{\frac{(N-1)\tau}{N}}(z)-\mathbf{I}N/\tau
\end{array}\right),
\label{Lforsto}
\end{equation}
where $\mathbf{I}$ is the identity matrix with dimension $\Omega$ and $ \mathcal{L}_{\frac{n\tau}{N}}$ is equivalent to
the tilted generator from Eq. \eqref{defL} with $\theta^n_{ij}$ instead of $\theta_{ij}(t)$. 

The maximal eigenvalue associated with \eqref{Lforsto} is written as  $\Lambda_N(z)$. 
We now show that $\lim_{N\to \infty}\Lambda_N(z)= \lambda(z)$. The right eigenvector associated with $\Lambda_N(z)$ is 
written as $\vec{v}_N(z)$. From the equation $\mathcal{L}(z)\vec{v}_N(z)=\Lambda_N(z)\vec{v}_N(z)$ and Eq. \eqref{Lforsto} we obtain
\begin{equation}
\frac{N}{\tau}\left[v_i^{n}(z)-v_i^{n-1}(z)\right]= \sum_{j}\left[\mathcal{L}_{\frac{n\tau}{N}}(z)\right]_{ij}v_j^{n}(z)-\Lambda_N(z)v_i^{n}(z), 
\end{equation}
where $v_i^{n}(z)$ are the components of $\vec{v}_N(z)$. We omit the dependence of these components on $N$ for a compact notation. 
This equation can be written in the form
\begin{equation}
\frac{N}{\tau}\left[\ket{v(z)}_n-\ket{v(z)}_{n-1}\right]= \mathcal{L}_{\frac{n\tau}{N}}(z)\ket{v(z)}_n-\Lambda_N(z)\ket{v(z)}_n,
\label{eqdisc} 
\end{equation}
where $\ket{v(z)}_n$ is vector with dimension $\Omega$ and components $v_i^{n}(z)$. In the limit $N\to\infty$, we set $n\tau/N\to t'$ and 
$\ket{v(z)}_n\to \ket{\overline{v}(z)}_{t'}$. Since for $n=N-1$ the stochastic protocol jumps back to 
$n=0$, by construction $\ket{\overline{v}(z)}_{t'}=\ket{\overline{v}(z)}_{t'+\tau}$. For $N\to\infty$, the vectorial form of Eq. \eqref{eqdisc} 
then becomes 
\begin{equation}
\frac{d}{dt'}\ket{\overline{v}(z)}_{t'}= \left[\mathcal{L}_{t'}(z)-\lim_{N\to\infty}\Lambda_N(z)\right]\ket{\overline{v}(z)}_{t'}.
\end{equation}
The formal solution of this equation reads 
\begin{equation}
\ket{\overline{v}(z)}_{t'}= \exp\left(-t'\lim_{N\to\infty}\Lambda_N(z)\right)\overleftarrow{\exp}\left(\int_0^{t'}\mathcal{L}_{u}(z)du\right) \ket{\overline{v}(z)}_{0}
\end{equation}
Using the periodicity of the eigenvector, i.e., $\ket{\overline{v}(z)}_{\tau}=\ket{\overline{v}(z)}_{0}$ and setting $t'=\tau$, we obtain
\begin{equation}
\mathcal{M}(z)\ket{\overline{v}(z)}_{\tau}= \exp\left(\tau\lim_{N\to\infty}\Lambda_N(z)\right)\ket{\overline{v}(z)}_{\tau},
\end{equation}
where $\mathcal{M}(z)$ is the fundamental matrix defined in Eq. \eqref{Mdef}. Since  by construction $\ket{\overline{v}(z)}_{\tau}$
is positive, from the Perron-Forbenius theorem 
$\exp\left(\tau\lim_{N\to\infty}\Lambda_N(z)\right)$ is the maximal eigenvalue associated with $\mathcal{M}(z)$, i.e., 
\begin{equation}
\lambda(z)=\lim_{N\to\infty}\Lambda_N(z).   
\end{equation}

\begin{figure}
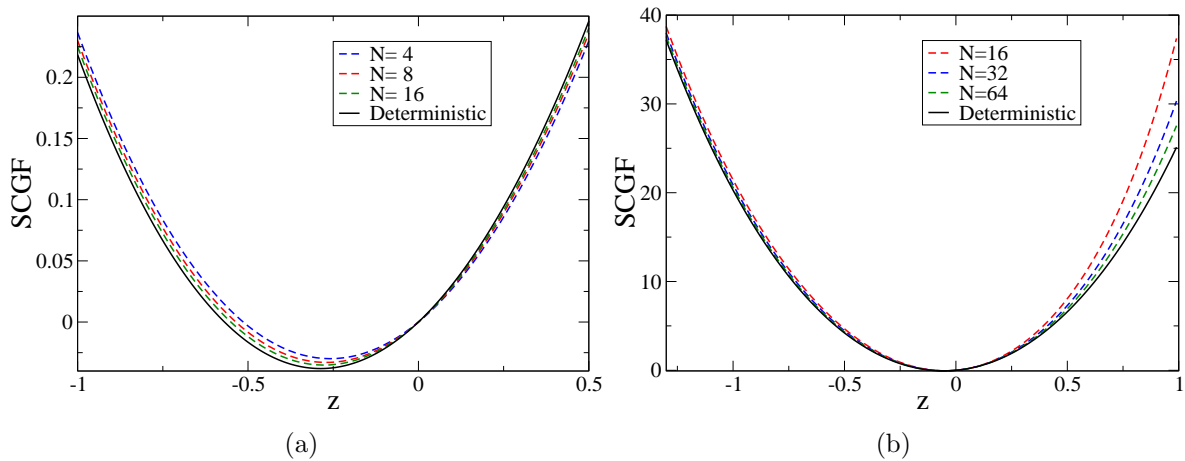

\subfigure[]{\includegraphics[width=78mm]{fig4a.eps}\label{fig4a}}
\subfigure[]{\includegraphics[width=75mm]{fig4b.eps}\label{fig4b}}
\vspace{-3mm}
\caption{Convergence of the SCGF associated with an stochastic protocol with $N$ jumps towards the SCGF associated with a deterministic protocol. (a)
Model for a molecular pump from Section \ref{sec42} with $\tau=w=1$. (b) Biased random walk with time-periodic affinity from Section \ref{sec43} with parameters $\psi=1$,
$\tau= 1$,  $w = 20$, and $F = 2$.}
\label{fig4}
\end{figure}

In Fig. \ref{fig4} we provide two numerical illustrations, one with the model from Section \ref{sec42} and the other with the model 
Section \ref{sec43}, of the convergence of 
the SCGF for a stochastic protocol with increasing $N$ towards the SCGF for a 
deterministic protocol. There is no generic inequality between $\Lambda_N(z)$ and $\Lambda(z)$.

\section{Discussion}
%%%%%%%%%%%%%%%%%%%%%%%%%%%%%%%%%%%%%%%%%%%%%%%%%%%%%%%%%%%%%%%%%%%%%%%%%%%%%%%%%%%%%%%%%%%%%%%%%%%%%%%%%%%%%%%%%%%%%%%%%%%%%%%%%%%%%%%%%%%%%%%%%%%%%%%%%
\label{sec6}

For systems driven by an external periodic protocol, which are well described by Markov processes with time-periodic transition rates, 
large fluctuations of a thermodynamic current can now be determined with the formalism developed in Section \ref{sec3}. In particular, 
the SCGF can be evaluated by calculating the maximum Floquet exponent associated with the fundamental matrix $\mathcal{M}(z)$.
Cumulants such as the average current $J$ and the diffusion coefficient $D$ can also be directly evaluated from this matrix, which provides 
a numerical method that can be more efficient than Monte Carlo simulations for small systems. Mathematically, beyond fluctuating currents, 
our formalism also applies to generic observables that count number of jumps with increments that do not have to be anti-symmetric.

We have calculated analytically the SCGF in two models: A heat engine with piecewise constant protocol and 
a biased random walk with a time-periodic affinity. Furthermore, we have verified our theoretical results  
by showing agreement between results obtained from numerical evaluation  
of the maximal Floquet exponent and results obtained from Monte Carlo simulations for a model for a molecular pump.

The SCGF associated  with a stochastic protocol with an infinitely large number of jumps is equivalent to the SCGF associated with a deterministic 
protocol, as proved in Section \ref{sec5}. This proof provides a rigorous basis to the idea that periodically driven systems can be analyzed with 
the use of stochastic protocols, i.e., if a result about current fluctuations is valid for stochastic protocols than it should be valid for deterministic protocols, 
which can be obtained as a particular limit of a stochastic protocol. The advantage of working with stochastic protocols is that the system and protocol together 
form a bipartite Markov process with constant transition rates that reach a nonequlibrium stationary state, which are quite well known.

\appendix

%--------------------------------------------------------------------------------------------------------------
\section{Work and related currents}
%--------------------------------------------------------------------------------------------------------------
\label{appnew}
A fluctuating current of interest in stochastic thermodynamics is the work done on the system 
due to the time variation of the energy levels. For jump processes, such current can be written in the form 
\begin{equation}
X'\left[A_{0}^{T}\right]\equiv\int_{0}^{T}f_{a_t}(t)dt,
\label{eqnewcurr}
\end{equation}
where  $f_{i}(t)= \partial_tg_i(t)$ and $g_i(t)$ is periodic with period $\tau$. For the case of 
work $g_i(t)$ is the free energy of state $i$. 

The empirical density $\rho_i(t)$ and the empirical flow $C_{ij}(t)$ are the number of periods for 
which the trajectory is in state $i$ at time $t\in [0,\tau]$ and the number of transitions from 
$i$ to $j$ at time $t$, respectively. They are functionals of the stochastic trajectory 
from time $0$ to time $T=n\tau$ that are defined as 
\begin{equation}
\rho_i(t)\equiv \frac{1}{n}\sum_{k=0}^n\delta_{a_{k\tau+t},i},
\label{eqnewrho}
\end{equation}
and
\begin{equation}
C_{ij}(t)\equiv \frac{1}{n}\sum_{k=0}^n\delta_{a_{(k\tau+t)^-},i}\delta_{a_{(k\tau+t)^+},j}.
\label{eqnewC}
\end{equation}
As shown in \cite{bert18}, in the large deviation regime, they fulfill the constraint
\begin{equation}
\frac{d}{dt} \rho(i,t)= \sum_j\left[C_{ji}(t)-C_{ij}(t)\right],
\label{meqfluc}
\end{equation}
for every single stochastic trajectory. Moreover, both $C_{ji}(t)$ and $\rho_{i}(t)$ are periodic with 
period $\tau$.  
  
Using the empirical density in Eq. \eqref{eqnewrho} we can rewrite Eq. \eqref{eqnewcurr} as 
\begin{equation}
X'\left[A_{0}^{T}\right]=n\sum_i\int_{0}^{\tau}\partial_tg_{i}(t)\rho_i(t)dt,
\label{eqnewcurr2}
\end{equation}
where we used $f_{i}(t)= \partial_tg_i(t)$. From the constraint in Eq. \eqref{meqfluc} 
and the definition of the empirical flow in Eq. \eqref{eqnewC}, we obtain  
\begin{equation}
X'\left[A_{0}^{T}\right]=\sum_{0\leq t\leq n\tau}\left(f_{t}(a_{t}^{+})-f_{t}(a_{t}^{-})\right),
\end{equation}
which is a current of the form given in Eq. \eqref{currtraj}. Hence, the current in Eq. \eqref{eqnewcurr} can be 
written as in Eq. \eqref{currtraj} if $f_{i}(t)= \partial_tg_i(t)$. This relation also holds for a stochastic 
protocol \cite{ray17}. In general, if $f_{i}(t)$ is not a derivative of a periodic function, this relation 
may not hold. Whereas extending the results from Sec. \ref{sec3} to such case should be straightforward, as we have done for 
diffusion processes in \ref{appa}, proving the equivalence with stochastic protocols for this more general current remains 
an open challenge.

%--------------------------------------------------------------------------------------------------------------
\section{Diffusion processes}
%--------------------------------------------------------------------------------------------------------------
\label{appa}
We consider a $d-$dimensional diffusion process $Y_t$ that follows the Fokker-Planck equation 
\begin{equation}
\partial_{t}\mu_{t}(y)=\mathcal{L}_{t}\left[\mu_{t}\right](y)
\label{eq:con-1}
\end{equation}
where $\mu_{t}(y)$ is the probability density at time $t$ and 
the adjoint of the generator is given by 
\begin{equation}
\mathcal{L}^{\dagger}_{t}=\widehat{F_{t}}\nabla+\nabla\frac{D_{t}}{2}\nabla.
\end{equation}
where $\widehat{F_{t}}$ is a time-periodic the drift vector, $D_{t}$ is a time-periodic diffusion matrix, $\tau$ is the period, and $\nabla$ is the nabla operator. 
This generator is then time-periodic. Hence, the invariant probability density related to the long time limit 
has the property $\mu^{\textrm{inv}}_{t}(y)=\mu^{\textrm{inv}}_{t+\tau}(y)$.

A stochastic current is a functional of the stochastic trajectory $Y_0^{n\tau}$, from time $0$ to time $n\tau$,
defined as 
\begin{equation}
X\left[Y_0^{n\tau}\right]\equiv\int_{0}^{n\tau}\left(f_{t}\left(Y_{t}\right)dt+g_{t}\left(Y_{t}\right)\circ dY_{t}\right),
\label{eq:J-1}
\end{equation}
where we use $\circ$ for the Stratonovich convention, $f_{t}=f_{t+\tau}$ is a scalar function, and $g_{t}=g_{t+\tau}$ is a vector field.
The Fokker-Planck current associated with $\rho_{t}^{\textrm{inv}}$ is a vector  
\begin{equation}
J_{\rho_{t}^{\textrm{inv}},t}(y)\equiv\widehat{F_{t}}(y)\rho_{t}^{\textrm{inv}}(y)-\frac{1}{2}D_t(y)\left(\nabla\rho_{t}^{\textrm{inv}}\right)(y).
\end{equation}
The typical behavior of $X\left[Y_0^{n\tau}\right]$ is given by 
\begin{equation}
\lim_{n\rightarrow\infty}\frac{X\left[Y_0^{n\tau}\right]}{n\tau}=\frac{1}{\tau}\int_{0}^{\tau}dt\left(\int\rho_{t}^{\textrm{inv}}(y)f_{t}(y)dy +\int J_{\rho_{t}^{\textrm{inv}},t}(y)g_{t}(y)dy\right).
\label{eq:tb-1}
\end{equation}

The SCGF is defined as 
\begin{equation}
\lambda(z)\equiv\lim_{n\rightarrow\infty}\frac{1}{n\tau}\ln\left(\left\langle \exp\left[ zX_{n\tau}\right]\right\rangle \right).
\label{eq:scg-2}
\end{equation}
Using the Feyman-Kac formula and the Grinasov lemma it can be shown that \cite{chet15} 
\begin{equation}
\langle\textrm{e}^{zX}\delta(Y_T-y')|Y_0=y\rangle= \left[\overleftarrow{\exp}\left(\int_{0}^{n\tau}dt\mathcal{L}_{t}(z)\right)\right](y',y).
\label{eq:fk-2}
\end{equation}
where the tilted generator $\mathcal{L}_{t}(z)$ is the second order differential operator of the form
\begin{equation}
\mathcal{L}^{\dagger}_{t}(z)=zf_{t}+\widehat{F_{t}}\left(\nabla+zg_{t}\right)+\left(\nabla+zg_{t}\right)\frac{D_{t}}{2}\left(\nabla+zg_{t}\right).
\end{equation}
Eq. \eqref{eq:fk-2} is analogous to Eq. \eqref{Mzpositive} for jump processes. The fundamental operator is defined as 
\begin{equation}
\mathcal{M}(z)\equiv\overleftarrow{\exp}\left(\int_{0}^{\tau}dt\mathcal{L}_{t}(z)\right).
\end{equation}
From Eq. \eqref{eq:fk-2}, this operator is also a Perron-Forbenius operator. Similar to the case for jump processes, $\mathcal{M}(z)$ can 
be expanded in the form given by Eq. \eqref{mexp}. The real maximum eigenvalue of $\mathcal{M}(z)$ is denoted $\rho_1(z)$. Following the same 
procedure for a jump process one can define a Laplace transform of the probability of current, analogous to Eq. \eqref{eqlaplace} (with an integral instead 
of a sum). The time evolution of this Laplace transform follows Eq. \eqref{laplaceevo}. Hence, the SCGF $\lambda(z)$ can be expressed in terms of 
the maximum eigenvalue $\rho_1(z)$ trough the formula \eqref{eqmainresult}.

%--------------------------------------------------------------------------------------------------------------
\section{Exact expressions for two-state model with piecewise constant protocol}
%--------------------------------------------------------------------------------------------------------------
\label{appb}
We consider a generic two-state model with a piecewise constant protocol (see \cite{verl13} for similar 
calculations for a protocol with two pieces). The transition 
rates during the interval $\tau_k$ are denoted by $w_+^{k}$ and $w_-^{k}$, where $w_+^k$ is 
the transition rate from state $1$ to state $2$. The increment of 
the current from state $1$ to state $2$  during the interval $\tau_k$ is $\theta^k$ and the increment for 
the reversed transition is $-\theta^k$. The piecewise version of the modified generator reads  
\begin{equation}
\tilde{\mathcal{L}}_{k}(z)=\left(\begin{array}{cc}
-w_+^{k} & w_-^{k}\textrm{e}^{-z\theta^k}\\
w_+^{k}\textrm{e}^{z\theta^k} & -w_-^{k}
\end{array}\right).
\end{equation}
From this matrix, we obtain 
\begin{equation}
\exp\left(-\tau_{k}\tilde{\mathcal{L}}_{n}(z)\right)=\frac{1}{w^k}
\left(\begin{array}{cc}
 \textrm{e}^{-\tau_{k}w^k}w_+^k+w_-^k  &  \textrm{e}^{-z\theta^k}(1-\textrm{e}^{-\tau_{k}w^k})w_-^k\\
\textrm{e}^{z\theta^k}(1-\textrm{e}^{-\tau_{k}w^k})w_+^k  &  w_+^k+\textrm{e}^{-\tau_{k}w^k}w_-^k
\end{array}\right)
\label{eq:brutal}
\end{equation}
where $w^k\equiv w_-^k+w_+^k$.
The fundamental matrix for a piecewise protocol in Eq. \eqref{Mpiece} is a product of the matrices in Eq. \eqref{eq:brutal}. 
Expressing the maximum eigenvalue of the two by two fundamental matrix in terms of its trace and determinant, and 
using the  Abel-Jacobi-Liouville identity for the determinant
\begin{equation}
\det\left(\mathcal{M}(z)\right)= \exp(\int_0^\tau Tr\left(\mathcal{L}_{t}(z)\right) dt)=\prod_{n=0}^{N-1}\textrm{e}^{-\tau_k w^k},
\end{equation}
we obtain the following expression for the scaled cumulant generating function 
\begin{equation}
\lambda(z)=\frac{1}{\tau}\ln\left[\frac{Tr\left(\mathcal{M}(z)\right)+\sqrt{\left[Tr\left(\mathcal{M}(z)\right)\right]^{2}-4\prod_{n=0}^{N-1}\textrm{e}^{-\tau_k w^k}}}{2}\right].
\end{equation}
Hence, we can calculate $\lambda(z)$ from the trace of the fundamental matrix in Eq. \eqref{Mpiece} with Eq. \eqref{eq:brutal}.
In particular, for $N=2$  the trace of the fundamental matrix is given by  
\begin{align}
Tr\left(\mathcal{M}(z)\right)  = & 4\textrm{e}^{-(\tau_0 w^0+\tau_1 w^1)/2}\sinh(w^0\tau_0/2)\sinh(w^1\tau_1/2)[q_+^1q_-^0\textrm{e}^{z(\theta^1-\theta^0)}+q_-^1q_+^0\textrm{e}^{z(\theta^0-\theta^1)}]\nonumber\\
&+(\textrm{e}^{-\tau_0 w^0}q_+^0+q_-^0)(\textrm{e}^{-\tau_1 w^1}q_+^1+q_-^1)+(q_+^0+\textrm{e}^{-\tau_0 w^0}q_-^0)(q_+^1+q_-^1\textrm{e}^{-\tau_1 w^1}),
\end{align}
where $q_\pm^0\equiv w_\pm^0/w^0$ and $q_\pm^1\equiv w_\pm^1/w^1$.

For the model for a heat engine in Section \ref{sec41}  the protocol has $N=4$ pieces. Since the changes in temperature are instantaneous, this number 
is reduced to $N=2$, with $\tau_0=\tau_1=\tau/2$. The transition rates for this model $w_+^0=w\textrm{e}^{-\beta_c E/2}$,  $w_-^0=w\textrm{e}^{\beta_c E/2}$,
$w_+^1=w\textrm{e}^{-\beta_h(E+\Delta E)/2}$, and $w_-^1=w\textrm{e}^{\beta_h (E+\Delta E)/2}$. For the current $X_q$ the increments,
are   $\theta_q^0=0$ and $\theta_q^1=\beta_c(E+\Delta E)$. For the current $X_e$ the increments are $\theta_e^0=0$ and $\theta_e^1=-1$.

\end{document}